\documentstyle[preprint,aps,psfig]{revtex}

\def\beq{\begin{equation}}
\def\eeq{\end{equation}}
\def\beqa{\begin{eqnarray}}
\def\eeqa{\end{eqnarray}}
\def\nn{\nonumber}
\def\al{&&}

\def\pa{\partial}
\def\square{\Delta}

\begin{document}

\begin{titlepage}

\vskip 40pt

\begin{flushright}

SISSA REF 5/98/EP

\end{flushright}

\vskip 40pt

\begin{center}

{\Huge Electric and magnetic interaction of dyonic\\

\vskip 5pt

D-branes and odd spin structure}
 
\vskip 20pt

{\large Matteo Bertolini, Roberto Iengo and Claudio A. Scrucca}

\vskip 20pt

International School for Advanced Studies ISAS-SISSA and INFN \\ 
Sezione di Trieste, Via Beirut 2-4, 34013 Trieste, Italy\\

\vskip 20pt
 
\end{center} 
 
\begin{abstract}

We present a general description of electromagnetic RR interactions between
pairs of magnetically dual D-branes, focusing on the interaction of a 
magnetically charged brane with an electrically charged one.
In the boundary state formalism, it turns out that while the electric-electric
and/or magnetic-magnetic interaction corresponds to the usual RR even spin 
structure, the magnetic-electric interaction is described by the RR odd 
spin structure.
As representative of the generic case of a dual pair of p and 6-p branes, 
we discuss in detail the case of the self-dual 3-brane wrapped on a 
$T_6/Z_3$, which looks like an extremal dyonic black hole in 4 dimensions.

\end{abstract}
 
\vskip 40pt

\begin{flushleft}



\end{flushleft}

\end{titlepage}

\section{Introduction}

Since Polchinski provided us with a powerfull $\sigma$-model technique for studing
non perturbative phenomena in String theory \cite{pol}, a number of interesting relations 
between String theory, Supergravity and Super Yang-mills theory have been understood.
In particular, the already known family of solitonic p-brane solutions of Type IIA and IIB
supergravities in $10$ dimensions \cite{stelle} are recognized to be described, in the String 
theory framework, by D-branes.  

A number of fascinating issues like black holes entropy and non-perturbative properties of 
Super Yang-Mills theory in diverse dimensions have been adressed in this context and partially 
answered, promoting D-branes and their dynamics \cite{bachas,kp} to one of the most promising and 
interesting parts of String theory to be studied. In particular, using the boundary state formalism 
\cite{polcai}, many properties of D-branes have been efficiently studied both in the covariant 
\cite{boun1,hins1,proc,divec} and in the light-cone \cite{boun2} formalisms.

In this paper we are going to study some interesting aspects of the electromagnetic
interactions between pairs of dual D-branes, which corresponds to the RR configuration
of the exchanged closed superstring. In the boundary state formalism, one has to 
further consider the two possible GSO signs, referred to as even and odd spin structure 
respectively. 
It is well known that the even spin structure encodes the standard interaction between two both 
electrically or both magnetically charged objects. We call this the diagonal interaction. An 
electrically charged object can also interact with a magnetically charged one. This interaction 
is more difficult to describe because the gauge potential fields cannot be globally defined. We 
call it the off-diagonal interaction, which also occurs in the general case of two dyonic objects
\cite{dyons}, carrying each both electric and magnetic charge (beside it, they will also have the 
diagonal interaction, of course). A general theoretical framework for describing the off-diagonal 
interaction has been developed long ago in ref. \cite{jecal}. In Sect. II, we review shortly this 
general framework, which is in fact very well suited for discussing the brane's interactions,
and we will show that some recently derived results for dyons in various dimensions \cite{dyons2}
are naturally obtained within this scheme.

In Sect. III we show that the general results of Sect. II for the off-diagonal interaction 
are exactly reproduced in String theory within the boundary state formalism, by an expression
of the amplitude in
the RR configuration corresponding to the odd spin structure. According to the general framework,
one has to consider the off-diagonal interaction of, say, one brane with a pair of a brane and an 
antibrane (it is like having a Dirac string between the two members of the pair).
In Sect. IV we consider in particular the interesting case of the D3-brane of the type IIB theory, 
which is self-dual in $d=10$ dimensions, that is both electrically and magnetically charged with 
respect to the self-dual RR 4-form present in the massless spectrum of the theory. We evaluate 
explicitly their diagonal and off-diagonal interactions. In Sect. V, we consider a wrapped 3-brane 
\cite{hins1,proc} in the interesting compactification over the $T_6/Z_3$ orbifold 
\cite{min}, leading in $4$ dimensions to an $N=2$ effective Supergravity theory \cite{n2}. 
We show how the single electric and magnetic charge in ten dimensions is reinterpreted from the 
4-dimensional non-compact spacetime point of view as a variety of possible dyonic charges, all 
satisfying Dirac's quantization condition, depending on the orientation of the brane in the compact
space. It is rather amusing to see how the odd spin structure string computation automatically 
encodes this feature. 

Let us end this introduction by remembering that, from the analysis of refs. 
\cite{hins1,proc}, the 3-brane on the $T_6/Z_3$ orbifold does not couple to 
(4-dimensional) scalars; rather, it only couples to gravity and to the $U(1)$ gauge field with 
equal strength, the total static diagonal interaction being zero, as appropriate for BPS states. 
Thus, being a source of equal strengh for gravity and Maxwell fields, and nothing else, it looks 
like a Reissner-Nordstr\"om black hole in 4-dimensional spacetime.

\section{Interactions of charges, monopoles and dyons}

As well kown, the electromagnetic potential generated by a magnetic monopole cannot be
defined everywhere; in the case of a $p$-extended object in $d$ spacetime
dimensions, there exists a Dirac hyperstring on which the potential is singular. 
As a consequence, the phase shift of another electrically charged $q$-dimensional 
extended object along a closed trajectory in this monopole background, which would be
a gauge-invariant quantity if the potential were well defined, suffers from an 
ambiguity. In fact, the requirement that the phase-shift should remain unchanged
mod $2 \pi$ leads to the famous Dirac quantization condition $eg = 2 \pi n$.

It is possible to define a mod $2 \pi$ gauge-invariant phase shift also for open trajectories 
by considering a pair of charge and anti-charge instead of a single charge.
Since an anti-charge travelling forward in time is equivalent to a charge travelling 
backward, this system can in fact be considered as a single charge describing a 
closed trajectory
\footnote{
If one consider only the usual electric-electric part of the interaction, one can 
even consider a single infinite straight trajectory; the corresponding phase-shift is 
gauge-invariant provided we require any gauge transformation to vanish at infinity.}.
The phase-shift for such a setting in the monopole background 
is then a gauge-invariant quantity (provided Dirac's quantization condition holds). 
Actually, this is the setting that can be most easily analyzed in the String theory 
framework, since it corresponds to D-branes moving with constant relative velocities. 
Indeed the available techniques for computing explicitely  branes interactions
allow us to deal only with rectilinear trajectories, more in general with hyperplanes 
as world surfaces.

The phase-shift for a system of a charge and an anti-charge moving along two parallel 
straight trajectories in a monopole background is a special case of the general analysis 
carried out in ref. \cite{jecal} that we shall briefly review.

We will consider dual pairs of branes, namely $p$-branes and $(d-4-p)$-branes (with  
$d$ being the dimension of the corresponding spacetime).
It is convenient to  describe the interactions formally in the Euclidean signature 
(which can be then continued to the Lorentz one). With such a metric one can consider 
closed world surfaces of the branes, as they would correspond, in Lorentz spacetime, 
to brane-antibrane pairs, as explained above. 

The world surface $\Sigma_{(p+1)}$ of the p-brane is $(p+1)$-dimensional 
and it couples to the $(p+1)$-form gauge potential $A_{(p+1)}$. We introduce the notation:
\beq
\int_{\Sigma_{(p+1)}} A_{(p+1)} \equiv \Sigma_{(p+1)}\cdot A_{(p+1)} \;.
\eeq
This can be rewritten as
\beq
\Sigma_{(p+1)} \cdot A_{(p+1)}=\Sigma_{(p+2)}\cdot F_{(p+2)} \;,
\eeq
where 
$F$ is the field strength $F_{(p+2)}=\nabla A_{(p+1)}$ and $\Sigma_{(p+2)}$
is an arbitrary $(p+2)$-dimensional surface whose boundary 
$\partial \Sigma_{(p+2)}$ is $\Sigma_{(p+1)}$. In formulae:
\beq
\Sigma_{(p+2)}\cdot \nabla A_{(p+1)}=\partial \Sigma_{(p+2)} \cdot A_{(p+1)}=
\Sigma_{(p+1)} \cdot A_{(p+1)} \;.
\eeq

The diagonal (electric-electric and/or magnetic-magnetic) interaction of two
$p$-branes, whose world surfaces are $\Sigma^\prime_{(p+1)}$ and 
$\Sigma_{(p+1)}$ respectively, can be written as
\beq
\label{d}
I_D=\left(e'e+g'g\right)\Sigma^\prime_{(p+2)} \cdot P \Sigma_{(p+2)}=
\left(e'e+g'g\right)\Sigma^\prime_{(p+1)} \cdot D \Sigma_{(p+1)} \;, 
\eeq
where $e,e'$ ($g,g'$) are the electric (magnetic) charges carried by the two 
branes, $D$ is the propagator, that is the inverse of the Laplace-Beltrami
operator $\square =\partial\nabla +\nabla\partial$, i.e. $\square D=1$, and 
$P = \nabla D \partial$. In the Euclidean path-integral, this interaction appears 
at the exponent, namely the integrand is $e^{-I_D}$.

Consider now what we call the off-diagonal interaction of two mutually dual
branes, a $p$-brane and a $(d-4-p)$-brane, in $d = 2(q+1)$ dimensions (the case
$p = q-1$ is self dual):
\beq
\label{offd2}
I_{off-D}= eg' \Sigma^\prime_{(d-2-p)} \cdot {^*P} \Sigma_{(p+2)}
+ e'g \Sigma_{(p+2)} \cdot {^*P} \Sigma^\prime_{(d-2-p)} \;.
\eeq
Here ${^*F} = \epsilon / 2^q F$ means the Hodge dual of a form $F$, obtained by 
contracting its components with the antisymmetric tensor. 
It is crucial to observe that the Hodge duality operation depends on the
dimension $d=2(q+1)$ of spacetime (that we shall suppose to be even in any case). 
In fact, the $\epsilon$ tensor satisfies $(\epsilon / 2^q)^2 = (-1)^{q+1}\,1$ and 
$\epsilon^T = (-1)^{q+1} \epsilon$.
Using these properties, one can see that $P + (-1)^{q+1} {^*P}\,{^*} = 1$ in the space of 
antisymmetric tensors, as it is equivalent to the Hodge decomposition.
Therefore ${^*P} + P\,{^*} = {^*}1$. Now, the insertion of the ${^*}1$ between 
$\Sigma^\prime_{(d-2-p)}$ and $\Sigma_{(p+2)}$ yields a contact term given by their
intersection number; assuming by a ``Dirac veto'' that this number is zero, we get 
${^*P} \doteq - P\,{^*}$. Finally, transposing the second term in eq. (\ref{offd2}) 
and using the above properties, we get finally
\beqa
\label{offd}
I_{off-D} &=& \left(eg' + (-1)^q e'g\right) 
\Sigma^\prime_{(d-2-p)} \cdot {^*P} \Sigma_{(p+2)} \nn \\ &=&  
\frac 1{2}\left(eg' + (-1)^q e'g\right) 
\left(\Sigma^\prime_{(d-2-p)} \cdot {^*P} \Sigma_{(p+2)}+(-1)^q \Sigma_{(p+2)} \cdot {^*P} 
\Sigma^{\prime}_{(d-2-p)}\right) \;.
\eeqa

In order for the path integral over $e^{iI_{off-D}}$ to be well defined, it is necessary
to impose the Dirac quantization condition \cite{dyons2}
\beq
\left(eg' + (-1)^q e'g\right) = 2 \pi n \;.
\eeq
The point is that $I_{off-D}$ depends on the (supposed irrelevant)
choice of the unphysical  $\Sigma^\prime_{(d-2-p)}$, which is only constrained
to have the physical brane world surface  $\Sigma^\prime_{(d-3-p)}$ as its
boundary:  $\pa\Sigma^\prime_{(d-2-p)}= \Sigma^\prime_{(d-3-p)}$.
However, the path-integral integrand is in this case $e^{iI_{off-D}}$
and this has no ambiguity. Indeed,
\beq
I_{off-D}=(2\pi n) \Sigma^\prime_{(d-2-p)} \cdot {^*\nabla} D \Sigma_{(p+1)} \;. 
\eeq
Now, if we change $\Sigma^\prime_{(d-2-p)}$ keeping its boundary fixed,
the ensuing change of $I_{off-D}$ can be written as 
$\delta I_{off-D}=(2\pi n)\pa {\cal V}_{(d-1-p)} \cdot {^*\nabla} D \Sigma_{(p+1)}$,
where the boundary of ${\cal V}_{(d-1-p)}$ is the union of the old
$\Sigma^\prime_{(d-2-p)}$ and the new one. By integrating by parts, using
${\nabla^*}={^*\pa}$ and $\pa \Sigma_{(p+1)}=0$ since we consider closed 
world surfaces, we get
\beq
\delta I_{off-D}=(2\pi n){\cal V}_{(d-1-p)}\cdot {^*\Sigma}_{(p+1)}
= 2\pi (\mbox{integer}) \;,
\eeq
since ${\cal V}_{(d-1-p)} \cdot {^*\Sigma}_{(p+1)}$ is the 
intersection number of the closed hypersurface $\Sigma_{(p+1)}$ and the 
hypervolume ${\cal V}_{(d-1-p)}$ and is therefore an integer.
Notice that relaxing the Dirac veto, eq. (\ref{offd}) is a consistent expression
provided $eg' + (-1)^q e'g = 4 \pi n$.

The above properties remain valid also when we compactify some of the
dimensions, in particular compactifying $6$ (the directions $a,a+1$, $a=4,6,8$) 
of the $d=10$ dimensions in String theory. 
Objects whose extended dimensions are wrapped in the compactified 
directions will appear point-like in the 4-dimensional spacetime. In particular, 
as anticipated, we will be interested in the sequel in the case of the $D3$-brane, 
occuring in Type IIB String theory, compactified on the orbifold $T_6/Z_3$. 
The 3-brane of Type IIB is a special case since it is both electrically and 
magnetically charged with respect to the self-dual RR 4-form; this peculiarity
will be relevant in our study giving rise, both before and after the 
compactification, to a dyonically charged state.
From the 4-dimensional spacetime point of view, this will look like the interaction 
of two dyons, whose values of electric and magnetic 
charges turn out to be dictated by the brane's different orientations in the compact 
directions. For instance, if the two (off-diagonally) interacting branes are parallel in 
the compact directions, then it is easy to see (we will be explicit in the 
following) that $I_{off-D}=(2\pi n) \Sigma^\prime_{(d-2-p)} \cdot {^*\nabla} D\Sigma_{(p+1)}=0$
and this will be interpreted in $4$ dimensions by saying that there is no 
off-diagonal interaction between to "parallel" dyons, that is having the same 
ratio (magnetic charge)/(electric charge). In fact, two such dyons behave with 
respect to each other as purely electrically charged particles.

It is amusing to notice that although the Dirac quantization condition 
is automatically implemented, as we said, once the off-diagonal interaction
is correctly normalized in $10$ dimensions, it might look somewhat 
non-obvious at first sight in $4$ dimensions, due to the non-intuitive
features of compact spaces. We will explore the ensuing pattern 
of charge quantization in the following sections.

\vspace{0.7cm}

In the following, we are going to consider the off-diagonal interaction of two pairs of 
3-branes-antibranes, wrapped on the compact space and moving linearly in spacetime 
(the brane's parameters will be labeled by $B$, the antibrane's ones by $A$ and the index 
$i=1,2$ labels the two pairs). We will take the trajectories in spacetime to 
describe a line in the $(t,x)$ plane. In each of the two pairs, the brane and the antibrane
are parallel to each other. This means that each pair is described by two parallel
$4$-dimensional hyperplanes, three directions being compact and specified by the angles 
$\theta^{(i)}_a$ ($a=4,6,8$), which are common to the brane and the antibrane, in each 
of the three tori which compose $T_6$ and one direction $w^{(i)}$ in the plane $(t,x)$. 
In the Lorentz spacetime, the $(t,x)$ direction $w^{(i)}$ is specified by an hyperbolic 
angle, the rapidity $v^{(i)}$ ($w_t^{(i)} = \sinh v^{(i)} \;,\;\; w_x^{(i)} = \cosh v^{(i)}$).
The $(t,x)$ trajectory of the brane of the pair $i$ is taken in the positive $t$-direction 
and is located at position $y^{(i)}_B,z^{(i)}_B$ in the transverse $(y,z)$ plane, while the 
trajectory of the antibrane is taken in the negative $t$-direction and is located at 
position $y^{(i)}_A, z^{(i)}_A$. 
It is convenient to introduce a complex variable $\xi = y + i z$. 
The positions of the brane and the antibrane of the two pairs in the transverse $(y,z)$ 
plane is depicted in Fig. 1.

\vskip 20pt

\input epsf
\epsfsize=50pt
\centerline{\epsffile{pos.eps}}
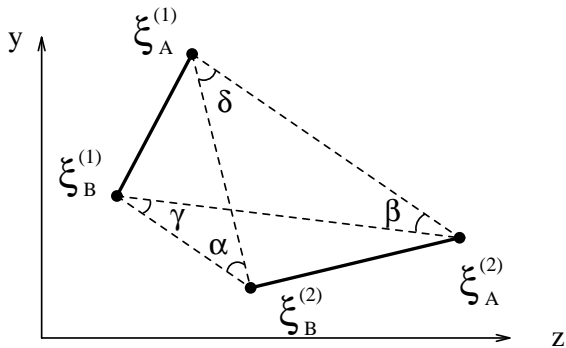
\centerline{\bf Fig. 1}

\vskip 10pt

According to the general construction, the diagonal and off-diagonal interactions
$I_{D}$ and $I_{off-D}$ are given by eqs. (\ref{d}) and (\ref{offd}) respectively.
In order to integrate along the hypersurfaces, let us suppose first that
the angles $\theta^{(2)}_a$  are different from the angles $\theta^{(1)}_a$. Consider then 
the Fourier transform of $D_d(r)=\int d^dk/(2\pi )^d \tilde D(k) e^{ikr}$ and
write $\tilde D(k)= 1/k^2 = \int_0^{\infty} dl e^{-lk^2}$.
The integration along the planes in the compact space and along the $(t,x)$ plane 
will result in putting to zero all the compact and the $(t,x)$ components 
of the momentum $k$. Hence, after those integrations, the propagator
$D$ will be reduced to the Fourier transform of $\tilde D$ where
only $k_y,k_z$ are different from zero, that is the
two dimensional propagator $D_2$ in the plane $(y,z)$. Thus, the only possible
derivatives occurring in the previous equation will be in the $(y,z)$ plane. 
Actually, by doing the integration over $l$ as the last one,
the other integrations factorize into the product of integrations
along the planes $(t,x)$, $(y,z)$ and the three compact planes $(a,a+1)$ 
respectively. 

In the diagonal case, the integration in the $(t,x)$ plane gives
$$
(w^{(1)} \cdot w^{(2)}) \int dt^{(1)} \int dt^{(2)} \int \frac {dk_tdk_x}{(2\pi )^2}
e^{i(t^{(1)}w^{(1)}-t^{(2)}w^{(2)})\cdot k} e^{-l(k_t^2+k_x^2)}=
\frac{w^{(1)}\cdot w^{(2)}}{|w^{(1)}\wedge w^{(2)}|} = \coth (v_1 - v_2)
$$
where  $w^{(i)}$ represents the direction of the $i$ branes trajectories in the 
$(t,x)$ plane.
The integrations in the $(a,a+1)$ planes give instead, as we will see in Sect. V, 
$$
\frac {\prod_a L_a^{(1)}L_a^{(2)}}{\mbox{Vol}(T_6/Z_3)} 
\prod_a \cos{(\theta^{(1)}_a -\theta^{(2)}_a)} \;.
$$
This factor (times the 10-dimensional charges $e'e + g'g$) is interpreted in 
4-dimensions as the dyon charge combination $e^{(1)}e^{(2)}+g^{(1)}g^{(2)}$.
It is convenient to introduce the two-dimensional {\it complex} propagator, whose
real part is $D_2 (\xi, \xi^\prime) = \mbox{Re} {\cal D}_2 (\xi, \xi^\prime)$
($\lambda$ is an infrared cut-off) 
\beq
{\cal D}_2(\xi, \xi^\prime) = \frac 1{2 \pi} \ln \frac {\xi - \xi^\prime}{\lambda} \;.
\eeq
The remaining integrations in the $(y,z)$ plane are over the straight lines joining 
the brane in $\xi^{(i)}_B$ and the antibrane in $\xi^{(i)}_A$ for each of the two 
pairs $i=1,2$, and give
\beqa
&&\left(e^{(1)}e^{(2)}+g^{(1)}g^{(2)}\right)
\int_{\xi^{(1)}_B}^{\xi^{(1)}_A} d \xi^{(1)} \cdot \pa_{\xi^{(1)}} 
\int_{\xi^{(2)}_B}^{\xi^{(2)}_A} d \xi^{(2)} \cdot \pa_{\xi^{(2)}} 
\mbox{Re} {\cal D}_2 (\xi^{(1)}, \xi^{(2)}) = 
\nn \\ && \qquad = \frac {\left(e^{(1)}e^{(2)}+g^{(1)}g^{(2)}\right)}{2 \pi}
\mbox{Re} \ln \left(\frac {\xi^{(1)}_A - \xi^{(2)}_A}{\xi^{(1)}_B - \xi^{(2)}_A} 
\cdot \frac {\xi^{(1)}_B - \xi^{(2)}_B}{\xi^{(1)}_A - \xi^{(2)}_B} \right) \;.
\eeqa 

In the off-diagonal case, the integration in the $(t,x)$ plane gives
$$
(w^{(1)}\wedge w^{(2)}) \int dt^{(1)} \int dt^{(2)} \int \frac {dk_tdk_x}{(2\pi )^2}
e^{i(t^{(1)}w^{(1)}-t^{(2)}w^{(2)})\cdot k} e^{-l(k_t^2+k_x^2)}=
\frac{w^{(1)}\wedge w^{(2)}}{|w^{(1)}\wedge w^{(2)}|} = \pm 1 \;.
$$
The result is therefore $\pm 1$ (the degenerate case where the trajectories 
$(1)$ and $(2)$ are parallel should be taken to be zero).
The integrations in the $(a,a+1)$ planes give instead
$$
\frac {\prod_a L_a^{(1)}L_a^{(2)}}{\mbox{Vol}(T_6/Z_3)} 
\prod_a \sin{(\theta^{(1)}_a -\theta^{(2)}_a)} \;.
$$
This factor (times the 10-dimensional charges $eg'+e'g$) is interpreted in 4-dimensions 
as the dyon charge combination $e^{(1)}g^{(2)}-g^{(1)}e^{(2)} = 2 \pi n$.
The remaining integrations in the $(y,z)$ plane give in this case (for $n=1$)
\beqa
&& \left(e^{(1)}g^{(2)}-g^{(1)}e^{(2)}\right) 
\int_{\xi^{(1)}_B}^{\xi^{(1)}_A} d \xi^{(1)} \wedge \pa_{\xi^{(1)}} 
\int_{\xi^{(2)}_B}^{\xi^{(2)}_A} d \xi^{(2)} \cdot \pa_{\xi^{(2)}} 
\mbox{Re} {\cal D}_2 (\xi^{(1)}, \xi^{(2)}) = 
\nn \\ && \qquad =
\mbox{Im} \ln \left(\frac {\xi^{(1)}_A - \xi^{(2)}_A}{\xi^{(1)}_B - \xi^{(2)}_A} 
\cdot \frac {\xi^{(1)}_B - \xi^{(2)}_B}{\xi^{(1)}_A - \xi^{(2)}_B} \right) 
\nn \\ && \qquad = \beta - \alpha = \delta - \gamma
\eeqa
(keeping the same sign convention for the angles, see Fig. 1).

There are here two important observation that we can make. First, considering
pairs of branes-antibranes automatically eliminates any infrared divergence. Second,
the off-diagonal interaction is given by the difference of the angles by which any 
curve joining $\xi^{(1)}_B$ and $\xi^{(1)}_A$ is seen from $\xi^{(1)}_B$ and $\xi^{(1)}_A$,
or viceversa. We thus see explicitely that $I_{off-D}$ is defined modulo $2\pi$.
Concluding, the total diagonal and off-diagonal interactions are given by
\beqa
I_{D} &=& \frac {\left(e^{(1)}e^{(2)}+g^{(1)}g^{(2)}\right)}
{\tanh (v^{(1)} - v^{(2)})} \mbox{Re} 
{\cal D}_2 \;, \\ I_{off-D} &=& \pm \left(e^{(1)}g^{(2)}-g^{(1)}e^{(2)}\right) 
\mbox{Im} {\cal D}_2 \;,
\eeqa
with 
$$
{\cal D}_2 = \ln \left(\frac {\xi^{(1)}_A - \xi^{(2)}_A}{\xi^{(1)}_B - \xi^{(2)}_A} 
\cdot \frac {\xi^{(1)}_B - \xi^{(2)}_B}{\xi^{(1)}_A - \xi^{(2)}_B} \right) \;.
$$

Notice the interesting fact that in $d=2(q+1)=10$, where the gauge field is a $q=4$ 
even form, the 3-brane is a dyon in the sense that it has $e = g = \mu_3 = \sqrt{2 \pi}$ and
that it has both a diagonal and an off-diagonal interaction with itself. 
In fact, the off-diagonal interaction is in this case proportional to 
$e^{(1)} g^{(2)} + e^{(2)} g^{(1)}$ (whereas for q odd it is proportional to 
$e^{(1)} g^{(2)} - e^{(2)} g^{(1)}$) and different from zero also for 
$e^{(1)} = e^{(2)}$, $g^{(1)} = g^{(2)}$. 
On the contrary, for $d=2(q+1)=4$, where the gauge field is a $q=1$ odd form, two 
``parallel'' dyons having $e^{(1)} = e^{(2)}$ and $g^{(1)} = g^{(2)}$ do not have any 
off-diagonal interaction, the latter beeing proportional to 
$e^{(1)} g^{(2)} - e^{(2)} g^{(1)}$.

It turns out from our analysis that the d=10 off-diagonal interaction, proportional to
$e_{10} g_{10}$, becomes automatically proportional to 
$e^{(1)}_{4} g^{(2)}_{4} - e^{(2)}_{4} g^{(1)}_{4}$ upon compactification down to d=4.
This happens because the off-diagonal interaction is proportional to the factor 
$\prod_a \sin (\theta_a^{(1)} - \theta_a^{(2)})$, which is zero when the branes 
$(1)$ and $(2)$ are seen by a non-compact observer to be parallel in the sense that 
$e^{(1)} = e^{(2)}$ and $g^{(1)} = g^{(2)}$. All of this will be explicitly shown in Sect. V.
More in general, notice that the off-diagonal interaction between two dyons $(1)$ and $(2)$
is symmetric both for $q$ even and for $q$ odd, under the exchange of every quantum number, 
$(1) \leftrightarrow (2)$. In fact, the transverse $(y,z)$ contribution to the amplitude,
that is ${\cal D}_2$, is symmetric, ${\cal D}_2 (1,2) = {\cal D}_2 (2,1)$, whereas
each pair of the remaining non-transverse directions $(t,x)$ and $(a,a+1)$ gives an 
antisymmetric contribution; therefore, since $e^{(1)} g^{(2)} + (-1)^q e^{(2)} g^{(1)}$ is 
symmetric for $q$ even and antisymmetric for $q$ odd, the total amplitude turns out to be 
symmetric in both cases (see eq. (\ref{offd})).

\section{The interactions in string theory}

As already noticed, the diagonal electric-electric and/or magnetic-magnetic
interaction between two p-branes is a well defined quantity also for open trajectories.
In this case, in fact, there is no strict necessity of considering interactions among 
pairs of brane-antibrane (although this is advisable to avoid infrared problems).
In string theory, the diagonal even interaction of just one brane at $\xi^{(1)}$ and 
one brane at $\xi^{(2)}$ is computed within the boundary state formalism to be
\cite{pol,bachas,kp,polcai,boun1,hins1,proc,divec}
\beq
{\cal A}_D =\frac{\mu_p^2}{16} \sum_{\alpha~even}
<v^{(1)},\theta^{(1)}_a,\xi^{(1)}|\int_0^{\infty} dl e^{-lH}
|v^{(2)},\theta^{(2)}_a,\xi^{(2)}>_{\alpha} \;,
\eeq
where $|...>$ is the boundary state representing the $p$-brane, $H$ is the closed superstring 
hamiltonian, $\mu_p$ is the RR charge 
of the p-brane and the factor $1/16$ comes from our conventional normalization. 
The even spin structure corresponding to the
case $\alpha=RR+$ (meaning the RR closed superstring sector and the GSO projection sign $=1$)
represents the diagonal electromagnetic interaction, whereas the two NSNS spin structures
$\alpha = NS\pm$ represent the gravitational one.
The main features of this diagonal amplitude are reviewed in Sect. IV. 

Let us stress here that only the even spin structure contributes. 
In fact, in the odd spin structure case, even if the rapidity tilt
$v^{(1)}-v^{(2)}$ and the angle tilt $\theta^{(1)}_a -\theta^{(2)}_a$ 
prevent the occurrence of fermionic zero modes in the directions
$(t,x)$ and $(a,a+1)$ with $a=4,6,8$, there still remain
fermionic zero modes in the $(y,z)$ transverse directions. The 
amplitude therefore vanishes since there is no insertion of 
operators to soak up those zero modes.

Now we show that also the off-diagonal interaction can be expressed
in String theory within the boundary state formalism. In this case,
as we have seen, it is necessary to consider at least the interaction of 
a brane-antibrane pair, say located at $\xi^{(1)}_{B,A}$, with one 
brane (or antibrane) located at $\xi^{(2)}$. According to the previous general
description, this interaction is expressed by an integral 
over a Dirac string joining $\xi^{(1)}_{B}$ and $\xi^{(1)}_{A}$,
which we represent parametrically by $\xi^{(1)}(s)$, $s=(0,1)$.
 
The expression of the off-diagonal odd amplitude is the following:
\beq
\label{oddstr}
{\cal A}_{off-D}=\frac{\mu_p^2}{16}\int_0^1 ds
<v^{(1)},\theta^{(1)}_a,\xi^{(1)}(s)|J(s)\bar J(s)\int_0^{\infty} dl e^{-lH}
|v^{(2)},\theta^{(2)}_a,\xi^{(2)}>_{RR-} \;,
\eeq
where the subscript  $RR-$ means that the braket is evaluated in the RR odd spin 
structure. Here $J,\bar J$ represent the left and right moving "supercurrents": 
$J=\pa X^{\mu}\psi_{\mu}$ and $\bar J=\bar\pa X^{\mu}\bar\psi_{\mu}$.
Along the Dirac string, $\pa, \bar \pa = \pa_s \mp i \pa_\tau$, where $\pa_\tau$
is the normal derivative, that is along the direction $\tau$ orthogonal to the Dirac string;
$\tau$ is therefore the (Euclidean) world sheet evolution time of the closed superstring.  

The odd spin structure case is now different from zero due to the 
supercurrent insertion. In fact since the odd amplitude vanishes unless 
there is the proper fermionic zero modes insertion, only the part of the 
insertion containing $\psi_y\bar\psi_z$ (or $z,y$ interchanged)
will contribute (for this reason the result would be the same also
inserting the complete supercurrent including also the ghost part). 
Since the boundary conditions essentially identify $\psi$ and
$\bar\psi$, we see that, due to the anticommuting properties of the
fermionic coordinates, 
$$
<(1)|J(s)\bar J(s)\int_0^{\infty} dl e^{-lH}|(2)>_{RR-}= 2i
<(1)|(\pa_s y\pa_{\tau} z-\pa_s z\pa_{\tau} y)
\int_0^{\infty} dl \psi^0_y\bar\psi^0_z e^{-lH}|(2)>_{RR-} \;.
$$
Now, in the odd spin structure case the contribution of the 
fermionic and bosonic oscillator modes is equal to 1, since the bosonic 
modes' contribution is exactly the inverse of the fermionic modes' one. 
Moreover, only the non-oscillator part of the inserted supercurrents contributes:
the fermions are necessarily zero modes as already explained and give 
an antisymmetric result; consequently, we are left with an antisymmetric
bosonic correlation which is zero except for the non-oscillator part.  
Thus it remains only the non-oscillator modes, both bosonic and fermionic, contribution.
The rest of the discussion now follows closely the general pattern
described in Sect. II. 
The part of that contribution from the coordinates directions
$(t,x)$ and $(a,a+1)$ with $a=4,6,8$ gives a position 
independent factor, which after compactification can be reinterpreted 
as the dyon charge combination $e^{(1)}g^{(2)}-e^{(2)}g^{(1)}$.
This will be explicitely discussed in Sect. V. It is interesting to
notice that the contribution of the fermionic non-oscillator modes is essential
in providing the correct "numerators" in the resulting expressions.

The position dependence of the amplitude comes from the $(y,z)$ non-oscillator modes contribution.
The fermionic zero modes' $(y,z)$ contribution, with our normalization, is equal to 1/2, 
due to the insertion of $\psi^0_y\bar\psi^0_z$. We further notice that for the bosonic
modes $ds(\pa_s y,\pa_s z)=(dy,dz)$ along the integration line, and that as an operator 
$(\pa_{\tau}y, \pa_{\tau}z)=-(\pa_y,\pa_z)$; therefore 
$ds (\pa_s y\pa_{\tau} z-\pa_s z\pa_{\tau} y) = dy\pa_z - dz\pa_y \equiv d \xi \wedge \pa_{\xi}$.
Moreover, for the transverse bosonic modes
$<\xi^{(1)}(s)|\int_0^{\infty} dl e^{-lH}|\xi^{(2)} >= D_2(\xi^{(1)}(s),\xi^{(2)})$, whereas the 
remaining non-transverse part of the amplitude gives $\pm i$.  
Finally we obtain 
\beq
\int_0^1 ds <(1)|J(s)\bar J(s)\int_0^{\infty} dl e^{-lH}|(2)>_{RR-}^{(y,z)}
=\int_{\xi^{(1)}_B}^{\xi^{(1)}_A} d \xi^{(1)} \wedge \pa_{\xi^{(1)}}
D_2(\xi^{(1)},\xi^{(2)}) \;,
\eeq
which reproduces precisely the expected result for the off-diagonal
interaction, see Sect. II.

As a final comment, one could also suspect that the odd spin structure 
contribution, eq. (\ref{oddstr}), might somehow automatically come from general 
world sheet supersymmetry considerations. In fact, it is known \cite{ver} that
the occurrence of the supercurrent insertion is dictated by the occurrence of 
the socalled supermoduli, which indeed are expected in the odd spin structure case.
Actually, in the cylinder case there is only one modulus, the previously
introduced $l$, and thus one would expect only one supermodulus and one 
supercurrent insertion. However in our case we are obliged to consider simoultaneously
the interaction of a brane and antibrane pair with a given brane 
(or antibrane). Thus it is not surprising to see the occurrence
of the pair of supercurrents $J$ and $\bar J$ as if the brane-antibrane
pair would entail the torus, rather than cylinder, topology.

Let us stress that in any case it is a fact that the boundary state amplitude
eq. (\ref{oddstr}) reproduces exactly the correct result for the off-diagonal 
electric-magnetic interaction.

\section{D3-branes in 10 Dimensions}

In this section, we make more explicit the content of the formulae of Sect. III by
briefly reviewing a series of results obtained in ref. \cite{hins1,proc} about the 
dynamics of D3-branes in $10$ dimensions, using the boundary state formalism. 
In particular we will consider the precise
structure of the amplitude for the scattering of two moving of such
D3-branes with an arbitrary orientation putting in evidence the various contributions 
coming from the four different spin structures arising in a closed string channel computation.

Let us start from a 3-brane configuration with Neumann boundary conditions in
the directions $t=X^0$ and $X^a$, and Dirichlet in $x=X^1,y=X^2,z=X^3$ and $X^{a+1}$, with 
$a=4,6,8$. The coordinates $X^a,X^{a+1}$ will eventually become compact. 
Consider then two of these 3-branes moving with 
velocities $V^{(1)} = \tanh v^{(1)}$, $V^{(2)} = \tanh v^{(2)}$ along the $1$ direction, at
transverse positions $\vec Y^{(1)}$, $\vec Y^{(2)}$, and tilted in $a,a+1$ planes with
generic angles $\theta^{(1)}_a$ and $\theta^{(2)}_a$.

The cylinder amplitude in the closed string channel is just a tree level 
propagation between the two boundary states, which are defined to implement the
boundary conditions defining the branes: 
\beq
\label{amp}
{\cal A}=\frac {\mu^2_3}{16}\int_{0}^{\infty}dl \sum_\alpha 
<v^{(1)},\theta_a^{(1)},\vec Y^{(1)}|e^{-lH}|v^{(2)},\theta_a^{(2)},\vec Y^{(2)}>_\alpha \;.
\eeq
As stated before, there are two sectors, RR and NSNS, corresponding to 
periodicity and antiperiodicity of the fermionic fields around the cylinder, 
and after the GSO projection there are four spin structures, $R\pm$ and 
$NS\pm$, corresponding to all the possible periodicities of the fermions 
on the covering torus. 

The configuration space boundary state can be written as the a product of delta 
functions enforcing the boundary conditions for the center of mass position operator 
$X_o^\mu$, that is a Fourier superposition of momentum states:
\beqa
|v,\theta_a,\vec Y>_B = 
&& \delta \left(\cosh v (X_o^1 - Y^1) - \sinh v X_o^0 \right) 
\delta \left(X_o^2 - Y^2 \right) \delta \left(X_o^3 - Y^3 \right) \nn \\
&&\prod_a \delta \left(\cos \theta_a (X_o^a - Y^a) 
+ \sin \theta_a X_o^{a+1} \right) |v,\theta_a> \nn \\
= && \int \frac{d^{6}\vec k}{(2\pi)^{6}}
e^{i \vec k_B \cdot \vec Y} |v,\theta_a> \otimes|k_B> \;,
\eeqa
with the boosted and rotated momentum
$$
k_B^\mu = (\sinh v k^1, \cosh v k^1, k^2, k^3, \cos \theta_a k^a, 
\sin \theta_a k^a) \;.
$$

Integrating over the momenta and taking into account momentum 
conservation which for non-vanishing $v \equiv v^{(1)} - v^{(2)}$ and 
$\theta_a \equiv \theta_a^{(1)} - \theta_a^{(2)}$
forces all the Dirichlet momenta but $k^2, k^3$ to be zero, the amplitude 
factorizes into a bosonic (B) and a fermionic (F) partition functions:
\beqa
\label{amp10}
{\cal A}&=&\frac {\mu_3^2}{2\sinh |v| \prod_a 2 \sin |\theta_a|} 
\int_{0}^{\infty}dl \int \frac {d k^2 d k^3}{(2\pi)^2} 
e^{i \vec k \cdot \vec b} e^{- k_B^2 l} \sum_\alpha Z_B Z^\alpha_F \nn \\
&=& \frac {\mu_3^2}{2\sinh |v| \prod_a 2\sin |\theta_a|} \int_{0}^{\infty}
\frac {dl}{4\pi l} e^{-\frac {b^2}{4 l}} \sum_\alpha Z_B Z^\alpha_F \;,
\eeqa
where $\mu_3 = \sqrt{2 \pi}$ is the 3-brane tension, 
$\vec b = \vec Y_T^{(1)} - \vec Y_T^{(2)}$ ($b = |\xi^{(1)} - \xi^{(2)}|$) is the 
transverse impact parameter (in the $2,3$ directions) and
$$
Z^\alpha_{B,F}=<v^{(1)},\theta_a^{(1)}|e^{-lH}|v^{(2)},\theta_a^{(2)}>^\alpha_{B,F} \;.
$$ 
In the above expression, only the oscillator modes of the string coordinates $X^\mu$
appear, since we have already integrated over the center of mass coordinate. Notice 
also that world-sheets with $l \ll b^2$ give a negligible 
contribution to the amplitude, and in the large distance limit 
$b \rightarrow \infty$ only world-sheets with $l \rightarrow \infty$ will contribute.

Notice finally that the amplitude ${\cal A}$ can be written, in agreement with the 
fact that it corresponds to a phase-shift, as a world sheet integral
\beq
{\cal A} = \mu_3^2 \int d\tau \prod_a \int d\xi_a \int_0^{\infty} dl (4 \pi l)^{-3} 
e^{- \frac {r^2}{4l}} \frac 1{16} \sum_\alpha Z_B Z^\alpha_F 
\eeq
in terms of the true distance
$$
r = \sqrt{\vec b^2 + \sinh^2 v \tau^2 + \sum_a \sin^2 \theta_a \xi_a^2} \;.
$$
In the limit $v, \theta_a \rightarrow 0$, translational invariance along 
the directions $1,a$ is restored and the integral over the world-sheet produces 
simply the volume $V_{3+1}$ of the 3-branes.

The remaining part of the boundary state has been explicitly constructed in 
ref. \cite{hins1} (see also \cite{proc}); after the GSO projection, the even part
of total partition function was found to be ($\eta(2il)$ being the Dedekin function)
\beqa
\al Z_B=\eta(2il)^4 \frac {2 i \sinh v}{\vartheta_1(i \frac v\pi|2il)} 
\prod_a \frac {2 \sin \theta_a}{\vartheta_1( \frac {\theta_a} \pi|2il)} \;, \\
\al Z_F^{even}=\eta(2il)^{-4}
\left\{\vartheta_2(i\frac{v}{\pi}|2il)\prod_a 
\vartheta_2( \frac {\theta_a} \pi|2il) 
\right. \nn \\ \al \qquad \qquad \qquad \qquad \left.
-\vartheta_3(i\frac{v}{\pi}|2il)\prod_a \vartheta_3( \frac {\theta_a} \pi|2il)
+\vartheta_4(i\frac{v}{\pi}|2il)\prod_a \vartheta_4( \frac {\theta_a} \pi|2il)
\right\} \;.
\eeqa
The even part of the amplitude represents the usual interplay of the RR
attraction and NSNS repulsion, leading to the well known BPS cancellation
of the interaction between two parallel D-branes (vanishing like $v^4$ for small
velocities).
In the large distance limit ($b,l \rightarrow \infty$), the behavior of the partition 
functions is
\beqa
\al Z_B \rightarrow 1 \;, \nn \\
\al Z_F^{even} \rightarrow 
2 \cosh v \prod_a 2 \cos \theta_a  
- 2 \left(2 \cosh 2v + \sum_a 2\cos 2 \theta_a \right) \;. \nn 
\eeqa

As we have seen in Sect. III, the odd part encodes instead the electric-magnetic 
off-diagonal RR interaction; due to the supercurrent insertion carrying the fermion fields 
$\psi^2,\psi^3$, it does not vanish, since the (2,3) fermionic zero modes are soaked up:
$$
Z_F^{odd}= \eta(2il)^{-4}
\vartheta_1(i\frac{v}{\pi}|2il)\prod_a \vartheta_1( \frac {\theta_a} \pi|2il) \;.
$$
Notice that in the odd spin structure, the oscillator's contribution cancel between
fermions and bosons by world sheet supersymmetry, and
$$
Z_B Z_F^{odd} = 2i \sinh v \prod_a 2\sin \theta_a \;.
$$
Remember also that the bosonic coordinates present in these supercurrents alter the non-oscillator 
part of the bosonic partition function precisely in the right way to allow the interpretation 
of Sect. II.

Summarizing, the diagonal interaction between two 3-branes at the positions $\xi^{(1)}$ and 
$\xi^{(2)}$ in the transverse (2,3) plane is, at large distances,
\beqa
\label{amp10even}
I_D = \mu_3^2 \coth v \prod_a \cot \theta_a D_2|\xi^{(1)} - \xi^{(2)}| \;,
\eeqa
where
$$
D_d(r) = \int \frac {d^d k}{(2 \pi)^d} \frac 
{e^{i \vec k \cdot \vec r}}{k^2} = \int_0^{\infty} dl 
\left(4 \pi l \right)^{-\frac d2} e^{-\frac{r^2}{4l}}
$$
is the Green function in $d$ dimensions.

The off-diagonal interaction between a 3-brane at transverse position $\xi^{(2)}$ and
a pair of 3-brane and antibrane at $\xi_B^{(1)}$ and $\xi_A^{(1)}$ is instead
the same all distances and given by
\beqa
\label{amp10odd}
I_{off-D} = \pm \mu_3^2 \int_{\xi_B^{(1)}}^{\xi_A^{(1)}} d \xi^{(1)} \wedge \pa_{\xi^{(1)}} 
D_2|\xi^{(1)} - \xi^{(2)}| \;.
\eeqa

\section{D3-brane on $T_6$ and $T_6/Z_3$}

In this section we shall apply the general construction that we have 
introduced to the case of the Type IIB 3-brane wrapped on the orbifold
$T_6/Z_3$. Compactifying the directions $a,a+1,a=4,6,8$ on $T_6$, one gets $N=8$ 
supersymmetry, which is further broken down to $N=2$ by the $Z_3$ moding, and this 
configuration was shown in ref. \cite{hins1} to correspond to a solution of the low
energy effective $N=2$ supergravity with no coupling to any scalar. 

$T_6/Z_3$ is the orbifold limit of a $CY$ manifold with Hogde numbers 
$h_{1,1} = 9$ and $h_{1,2} = 0$. The standard counting of hyper
and vector multplets for Type IIB compactifications tells us that  
$n_V = h_{1,2}$ and $n_H= h_{1,1} + 1$ \cite{vafalou}
and the 4-dimensional low energy effective theory we are left with is 
therefore $N=2$ supergravity coupled to 10 hypermultiplets and 0 vector 
multiplets (see \cite{n2} and references therein). 
In particular, the only vector field arising in the compactification, namely 
the graviphoton, comes from the self-dual RR 4-form $C_{\mu\nu\rho\sigma}$  
under which the D3-brane is already charged in 10 dimensions.

As explicitly shown in ref. \cite{hins1,proc}, the 3-brane wrapped on 
$T_6/Z_3$ does not couple to the hypers (as it must be) and has both an electric 
and a magnetic charge with respect to the graviphoton, consistently with the fact 
that the 3-brane is selfdual in 10 dimensions.
This can be seen by analyzing the velocity dependence of the large 
distance behavior of the scattering amplitude for two of these 3-branes moving
with constant velocities in the 4-dimensional non-compact spacetime, in which 
they look point-like. 
The boundary state decribing this 3-brane wrapped on $T_6/Z_3$ can be obtained
from the one constructed for the non-compact 3-brane essentially through the
usual quantization of the momentum along a compact direction.

More precisely, recall that the $T_6/Z_3$ orbifold is constructed identifying
points in the covering $T_6 = T_2 \times T_2 \times T_2$ which are connected 
by $Z_3$ rotations in the 3 $a,a+1$ planes corresponding to each of the 
$T_2$'s \cite{min}. Notice that $h_{1,2} = 0$ means that the number of complex 
deformations is $0$ in this case, consistently with the fact that the 
Z-orbifold procedure ``freezes out'' any possible freedom in the choice of the 
3 $T_2$'s \cite{vafalou}. This reflects into the fact that  the 3-brane 
configuration we consider {\it must} have one Neumann and one Dirichlet 
direction in each of the  3 $T_2$'s and is therefore wrapped on a 3-cycle which
is ``democratically'' embedded in $(T_2)^3$.

\vspace{0,7cm}

Let us start concentrating on a single $T_2$ factor, then. The only lattice compatible
with the eventual $Z_3$ moding is the triangular one, with modulus 
$\tau = R e^{i \frac {\pi}3}$, as in Fig. 2.
The lattice of windings $\bar L = L_x + i L_y$ is given by 
$\bar L = m \tau + n R = \frac {R}2 (2n + m) + i \frac {\sqrt{3}}2 R m$, with 
$m,n$ integers, that is
$$
L_x = \frac R2 N_x \;,\;\; L_y = \frac {\sqrt{3}}2 R N_y \;,
$$
where $N_x,N_y$ are integers of the same parity.
The lattice of momenta is as usual determined by the requirement that the plane
wave $e^{i p \cdot X}$ is well defined when $X$ is shifted by a vector belonging to
the winding lattice, and one finds
$$
p_x = \frac {2 \pi}{R} n_x \;,\;\; p_y = \frac {2 \pi}{\sqrt{3}R} n_y \;,
$$
where $n_x,n_y$ are again integers of the same parity.

\vskip 20pt

\input epsf
\epsfsize=50pt
\centerline{\epsffile{t2.eps}}
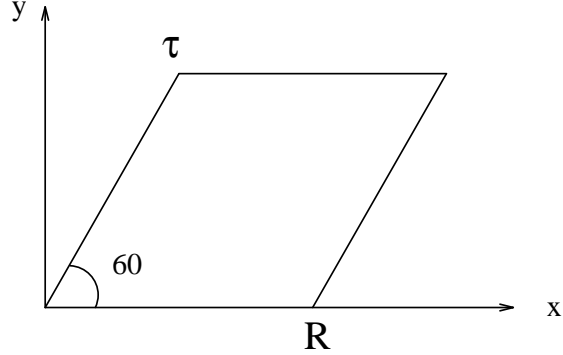
\centerline{\bf Fig. 2}

\vskip 10pt

We choose in each of the $T_2$ an arbitrary Dirichlet direction $x^\prime$ at 
angle $\theta$ with the $x$ direction and an orthogonal Neumann direction 
$y^\prime$ at angle $\Omega = \theta + \frac \pi 2$ with the $x$ direction, 
and fix its lenght. 
This amounts to choose an arbitrary vector $\bar L$ in the winding lattice, 
which is identified by the pair $(N_x,N_y)$ or, more conveniently for the
following, by the orthogonal pair $(\bar n_y, -\bar n_x)$, which corresponds 
to the orthogonal direction of allowed momenta (see Fig. 3).
In this way    
\beqa
&&L_x = - L \sin \theta \;,\;\; L_y = L \cos \theta \;, \nn \\
&&\cos\theta = - \frac {{\sqrt{3}} R}{2 L}\,\bar n_x\;,\;\;
\sin\theta = - \frac R{2 L}\,\bar n_y \;. \nn
\eeqa
where
$$
L \equiv \vert \bar L \vert 
= \frac R2 \sqrt{\bar n_y^2 + 3 \bar n_x^2} \;.
$$

\vskip 20pt

\input epsf
\epsfsize=100pt
\centerline{\epsffile{ang.eps}}
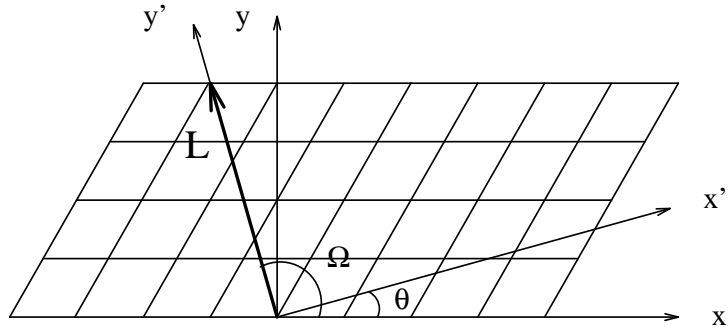
\centerline{\bf Fig. 3}

\vskip 10pt

We are now interested in the bosonic non oscillator modes contribution to the whole 
picture and let us start, for semplicity, remembering the result for the 
non-compact case. The boundary state for the bosonic non oscillator modes in a given $a,a+1$ 
pair is  
\beqa
\label{bcont}
|\vec Y>_B = && \delta \left(X'_o-Y'\right) |0> \nn \\
= && \int \frac{dp_x dp_y}{(2\pi)}
e^{-i (p_x \cdot Y_x + p_y \cdot Y_y)}\delta 
\left(\cos \theta p_y - \sin \theta p_x\right) |p_x,p_y> \;. 
\eeqa
The $\delta$-function selects momenta parallel to the Dirichelet direction we 
have chosen. Indeed if $\omega$ is the direction of the generic $\vec p$ momentum, 
the argument of the $\delta$-function becomes proportional to 
$\sin (\theta - \omega)$.
Using of the normalization
$$
<p^{(1)}_x, p^{(1)}_y|p^{(2)}_x, p^{(2)}_y> = (2\pi)^2 
\delta \left(p^{(1)}_x-p^{(2)}_x\right) 
\delta \left(p^{(1)}_y-p^{(2)}_y\right) \;,
$$
one recovers the following vacuum amplitude
\beqa
\label{ampc}
<\theta^{(1)},\vec Y^{(1)}|e^{-lH}|\theta^{(2)},\vec Y^{(2)}>_B =&& \int {dp_x dp_y}
e^{-i (p_x \cdot \Delta Y_x + p_y \cdot \Delta Y_y)} \times \nn \\
&& \times \delta \left(\cos \theta^{(1)} p_y - \sin \theta^{(1)} p_x\right)
\delta \left(\cos \theta^{(2)} p_y - \sin \theta^{(2)} p_x\right)  \nn \\
=&&\frac 1{\sin |\theta^{(1)} - \theta^{(2)}|} \;. 
\eeqa

In discretizing this result we adopt the following strategy. Let us begin by supposing 
$\theta^{(1)} \neq \theta^{(2)}$. First we substitute in eq. (\ref{ampc}) the previously 
derived  expressions for the discretized quantities $\vec p$ and $\theta$ and extract 
some jacobians from the Dirac $\delta$-functions, obtaining
\beqa
<\theta^{(1)},\vec Y^{(1)}|e^{-lH}|\theta^{(2)},\vec Y^{(2)}>_B &=& 
\frac{L(\theta^{(1)}) L(\theta^{(2)})}{(\sqrt 3/4)R^2}
\sum \hspace{-20pt}\raisebox{-12pt}
{$\scriptscriptstyle{n_x,n_y}$}
\hspace{-25pt}\raisebox{-20pt}
{$\scriptscriptstyle{\mbox{\tiny same par}}$}
\delta \left(\bar n^{(1)}_x n_y - \bar n^{(1)}_y n_x\right)
\delta \left(\bar n^{(2)}_x n_y - \bar n^{(2)}_y n_x\right)\;. \nn 
\eeqa
Since in this case the solution of the condition enforced by the $\delta$-functions 
is $n_x = n_y = 0$, all the momenta are zero and the exponential drops as in the 
continuum case.

The Dirac $\delta$-function containing only integers can now be turned to
a Kroneker one; however, since the latter is insensitive to an integer 
rescaling whereas the former transforms with an integer jacobian, we shall
keep an arbitrary integer constant in this step:
$$
\delta \left(\bar n^{(1)}_x n_y - \bar n^{(1)}_y n_x\right)
\delta \left(\bar n^{(2)}_x n_y - \bar n^{(2)}_y n_x\right)
= N \delta_{\bar n^{(1)}_x n_y, \bar n^{(1)}_y n_x}
\delta_{\bar n^{(2)}_x n_y, \bar n^{(2)}_y n_x}
= N \delta_{n_x, 0} \delta_{n_y, 0} \;.
$$ 
Therefore 
\footnote{
Notice that we consistently take
$\;
\sum \hspace{-20pt}\raisebox{-12pt}
{$\scriptscriptstyle{n_x,n_y}$}
\hspace{-25pt}\raisebox{-20pt}
{$\scriptscriptstyle{\mbox{\tiny same par}}$}
\delta_{n_x, 0} \delta_{n_y, 0} = \frac 12
$.}, with $\mbox{Vol}(T_2) = (\sqrt 3/2)R^2$
$$
<\theta^{(1)},\vec Y^{(1)}|e^{-lH}|\theta^{(2)},\vec Y^{(2)}>_B = 
N \frac{L(\theta^{(1)}) L(\theta^{(2)})}{\mbox{Vol}(T_2)}.
$$
The integer $N$ is fixed to $1$ by the requirement that for 
$\theta^{(1)} = \theta^{(2)}$ the amplitude reduces to the ``winding''
$L^2 / {\mbox{Vol}(T_2)}$. Actually, in order to achieve 
the above limit, an infinite $L(\theta)$ is in general required because
of the discreteness of the allowed angles, even if for strictly parallel branes
finite $L(\theta)$'s are possible.
Indeed, $L(\theta^{(1)}) L(\theta^{(2)})\sin|\theta^{(1)} - \theta^{(2)}|=
|\bar n^{(1)}_x \bar n^{(2)}_y - \bar n^{(1)}_y \bar n^{(2)}_x|\mbox{Vol}(T_2)$. 
In this way the continuum and discrete results differ by the integer jacobian 
$|\bar n^{(1)}_x \bar n^{(2)}_y - \bar n^{(1)}_y \bar n^{(2)}_x|$ (which vanishes
for $\theta^{(1)} = \theta^{(2)}$).
The final result is then
\beq
\label{ampd}
<\theta^{(1)},\vec Y_1|e^{-lH}|\theta^{(2)},\vec Y_2>_B =
\frac{L(\theta^{(1)}) L(\theta^{(2)})}{\mbox{Vol}(T_2)}=
\frac{|\bar n^{(1)}_x \bar n^{(2)}_y - \bar n^{(1)}_y \bar n^{(2)}_x|}
{\sin|\theta^{(1)} - \theta^{(2)}|} \;.
\eeq
The above result could have been obtained starting directly from the compact 
boundary state, that is, by {\it first} discretizing the continuum boundary state
(\ref{bcont}) and {\it then} computing the amplitude. 
The correct discrete boundary state turns out to be 
\beqa
|\vec Y>_B = L(\theta)\sum \hspace{-20pt}\raisebox{-12pt}
{$\scriptscriptstyle{n_x,n_y}$}
\hspace{-25pt}\raisebox{-20pt}
{$\scriptscriptstyle{\mbox{\tiny same par}}$} \frac1{(\sqrt 3/2) R^2}
e^{-\frac {2 \pi}R i (n_x Y_x + n_y / \sqrt 3 Y_y)}\delta \left
(\bar n_x n_y - \bar n_y n_x\right) |n_x,n_y> \;,
\eeqa 
and reproduces correctly eq. (\ref{ampd}) with the definition 
$$
<n_x,n_y|m_x,m_y> = \sqrt{3} R^2 \delta_{n_x,m_x} \delta_{n_y,m_y} \;.
$$

\vspace{0.7cm}

Postponing for the moment the $Z_3$ identification, let us now consider 
as an instructive intermediate result the case of $T_6$.
The result eq. (\ref{ampd}) can be generalized in a straightforward way 
giving for the total contribution from the compact part of the bosonic non oscillator modes
\beq
\label{boszer}
<\theta_a^{(1)},\vec Y^{(1)}|e^{-lH}|\theta_a^{(2)},\vec Y^{(2)}>_B = 
\frac{V(B_1) V(B_2)}{\mbox{Vol}(T_6)} \;, 
\eeq
where $V(B_1)$, $V(B_2)$ are the volumes of the two 3-branes. This factor can be 
reabsorbed in the definition of a 4-dimensional $\hat \mu_3$ (from now on 
$\theta_a^{(1)} - \theta_a^{(2)} \equiv \theta_a$) 
\beq
\label{mu}
\hat \mu_3^2 \equiv \mu_3^2 \frac{V(B_1) V(B_2)}{\mbox{Vol}(T_6)} = 2 \pi \prod_a 
\frac{|\bar n^{(1)}_a \bar n^{(2)}_{a+1} - \bar n^{(1)}_{a+1} \bar n^{(2)}_a|}
{\sin |\theta_a|} \;.
\eeq

The contribution of the fermions doesn't change during the compactification 
and the amplitude (\ref{amp10}) becomes in this case
\beqa
\label{amp4}
{\cal A} &=& \frac {\hat \mu_3^2}{\sinh |v|} 
\int_{0}^{\infty}
\frac {dl}{4\pi l} e^{-\frac {b^2}{4 l}} \frac 1{16} \sum_s Z_B Z^s_F \;,
\eeqa
and can be rewritten this time as a {\it one} dimensional world-sheet integral
\beq
{\cal A} = \hat \mu_3^2
\int d\tau \int dl (4 \pi l)^{-\frac 32} 
e^{- \frac {r^2}{4l}} \frac 1{16} \sum_s Z_B Z^s_F \;,
\eeq
in terms of the 4-dimensional distance
$$
r = \sqrt{\vec b^2 + \sinh^2 v \tau^2} \;.
$$

Eqs. (\ref{amp10even}) for the large distance diagonal interaction between two branes 
at the positions $\xi^{(1)}$ and $\xi^{(2)}$, and (\ref{amp10odd}) for the 
scale-independent off-diagonal interaction between a brane at transverse 
position $\xi^{(2)}$ and a pair of brane and antibrane at $\xi_B^{(1)}$ and 
$\xi_A^{(1)}$, modify to
\beqa
\label{amp4even}
I_D &=& \alpha_{even} \coth v D_2|\xi^{(1)} - \xi^{(2)}| \;, \\
\label{amp4odd}
I_{off-D} &=& \pm \alpha_{odd} \int_{\xi_B^{(1)}}^{\xi_A^{(1)}}
d \xi^{(1)} \wedge \pa_{\xi^{(1)}} D_2|\xi^{(1)} - \xi^{(2)}| \;,
\eeqa
with 
\beqa
\alpha_{even} &=& \hat \mu_3^2 \prod_a \cos \theta_a \;, \nn \\
\alpha_{odd} &=& \hat \mu_3^2 \prod_a \sin \theta_a \;.
\nn
\eeqa
Recalling (\ref{mu}) and noticing that
$$
\cot \theta_a = \sqrt 3 \frac
{3 \bar n^{(1)}_a \bar n^{(2)}_a + \bar n^{(1)}_{a+1}\bar n^{(2)}_{a+1}}
{\bar n^{(1)}_a \bar n^{(2)}_{a+1} - \bar n^{(1)}_{a+1}\bar n^{(2)}_a} \;,
$$
the two coupling can also be written as
\beqa
\label{torch}
\alpha_{even} &=& 2 \pi \prod_a \sqrt 3 
\left(3 \bar n^{(1)}_a \bar n^{(2)}_a + \bar n^{(1)}_{a+1}\bar n^{(2)}_{a+1}\right)
\;, \nn \\ 
\alpha_{odd} &=& 2 \pi \prod_a 
\left(\bar n^{(1)}_a \bar n^{(2)}_{a+1} - \bar n^{(1)}_{a+1}\bar n^{(2)}_a\right) \;.
\eeqa

As expected, the orientation of the 3-branes in 10 dimensions affects
the effective electric and magnetic couplings of the correspondig 0-branes in 
4 dimensions. 
Notice that the Dirac quantization condition for the off-diagonal coupling 
$\alpha_{odd}$, which is satisfied in 10 dimensions with the 
minimal allowed charges \cite{pol}, remains satisfied in 4 with an 
integer which depends on the branes' orientation.
This result can also be understood in terms of the relevant 
$N=8$ supergravity. Notice in fact that 
\beqa
\prod_a \cos \theta_a &=& \frac 14 \sum_{i=1}^4 \cos \phi_i \;, \nn \\
\prod_a \sin \theta_a &=& - \frac 14 \sum_{i=1}^4 \sin \phi_i \;, \nn
\eeqa
with $\phi_i \equiv \phi^{(1)}_i - \phi^{(2)}_i$ and
\beqa
\phi^{(1,2)}_1 = \theta^{(1,2)}_4 + \theta^{(1,2)}_6 + \theta^{(1,2)}_8 &&\;,\;\;
\phi^{(1,2)}_2 = - \theta^{(1,2)}_4 - \theta^{(1,2)}_6 + \theta^{(1,2)}_8\;, \nn \\
\phi^{(1,2)}_3 = \theta^{(1,2)}_4 - \theta^{(1,2)}_6 - \theta^{(1,2)}_8 &&\;,\;\;
\phi^{(1,2)}_4 = - \theta^{(1,2)}_4 + \theta^{(1,2)}_6 - \theta^{(1,2)}_8 \;. \nn
\eeqa
The effective couplings can thus be rewritten as
\beqa
\alpha_{even} &=& \sum_{i=1}^4 \left(e^{(1)}_i e^{(2)}_i + g^{(1)}_i g^{(2)}_i 
\right) \;, \nn \\
\alpha_{odd} &=& \sum_{i=1}^4 \left(e^{(1)}_i g^{(2)}_i - g^{(1)}_i e^{(2)}_i 
\right) \;,
\eeqa
with 
\beqa
&&e^{(1)}_i = \frac {\hat \mu_3}2 \cos \phi^{(1)}_i \;,\;\;
e^{(2)}_i = \frac {\hat \mu_3}2 \cos \phi^{(2)}_i \;, \nn \\
&&g^{(1)}_i = \frac {\hat \mu_3}2 \sin \phi^{(1)}_i \;,\;\;
g^{(2)}_i = \frac {\hat \mu_3}2 \sin \phi^{(2)}_i \;.
\eeqa

This second consideration allows to keep track of the coupling to the various 
vector fields. 
In fact it happens that the ten vectors fields arising from dimensional 
reduction of the RR 4-form, couple to the brane only through four independent 
combinations of fields, with electric and magnetic charges parametrized by the 
four angles $\phi^{(1,2)}_i$. Since the electric and magnetic charges correponding
to a given $\phi^{(1,2)}_i$ cannot vanish simultaneously, the $3$-brane cannot 
decouple from any of the four effective gauge fields, in agreement with a pure 
Supergravity argument achieved in ref. \cite{n8su}.
From this point of view, the Dirac quantization condition, emerging clearly in 
(\ref{torch}), is to be understood on the sum of the couplings corresponding to the
four independent $\phi^{(1,2)}_i$, and not on the charges with respect to the single 
fields.

The whole picture determines therefore a 4-parameter family of dyons which are 
inequivalent from the 4-dimensional point of view since they carry a different 
set of charges.
Notice finally that when two of these branes have equal $\phi^{(1,2)}_i$'s 
(yielding vanishing $\phi_i$'s) their diagonal coupling no longer depends on the 
angles and the off-diagonal one vanish, as appropriate for identical dyons in $d=4$ 
dimensions. 

\vspace{0.7cm}

Let us discuss finally the orbifold case. As explained in ref. \cite{hins1},
the only effect of the $Z_3$ moding is to project the boundary state for $T_6$
onto its $Z_3$-invariant part. This projection can be easily performed by first 
computing the amplitude on $T_6$ with a relative twist $z_a$ in the orientations,
$\theta_a \rightarrow \theta_a + 2 \pi z_a$, and then averaging finally on all the 
possible $z_a$'s 
\footnote{The twists $z_a$ in the 3 $a,a+1$ planes satisfy $\sum_a z_a = 0$ in
order to preserve at least one supersymmetry \cite{min}. 
The allowed sets $\{z_a\}$ of relative twits can be taken to be 
$\{(0,0,0), (1/3,1/3,-2/3), (2/3,2/3,-4/3)\}$.}.

Since the bosonic zero modes' contribution (\ref{boszer}) does not depend explictly 
on the angles, the only modification introduced by the $Z_3$ moding is in the 
volume: $\mbox{Vol}(T_6/Z_3) = 1/3 \mbox{Vol}(T_6)$. For the fermions, instead,
one simply sets $\theta_a \rightarrow \theta_a + 2 \pi z_a$; under this relative 
rotation one has correspondingly:
\beqa
&&\phi_1 \rightarrow \phi_1 + 2 \pi (z_4  + z_6 + z_8) = \phi_1 \;, \nn \\
&&\phi_2 \rightarrow \phi_2 + 2 \pi (-z_4  - z_6 + z_8) 
= \phi_2 + 4 \pi  z_8 \;, \nn \\
&&\phi_3 \rightarrow \phi_3 + 2 \pi (z_4  - z_6 - z_8) 
= \phi_3 - 4 \pi z_4 \;, \nn \\
&&\phi_4 \rightarrow \phi_4 + 2 \pi (-z_4  + z_6 - z_8) 
= \phi_4 + 4 \pi z_6 \;.\nn
\eeqa
The averaging procedure has the important consequence of projecting out the
contribution depending on the non invariant $\phi_2, \phi_3, \phi_4$,
with respect to the $T_6$ case. 
Indeed,
\beqa
\label{orbch}
\alpha_{even} = \hat \mu^2_3 \sum_{\{z_a\}} \prod_a \cos (\theta_a + 2 \pi z_a) 
= \frac {\hat \mu^2_3}{4} \cos \phi_1 \;, \nn \\
\alpha_{odd} = \hat \mu^2_3 \sum_{\{z_a\}} \prod_a \sin (\theta_a + 2 \pi z_a) = 
-\frac {\hat \mu^2_3}{4} \sin \phi_1 \;.
\eeqa
where the $1/3$ of the averiging has canceled with the $3$ coming from the volume of 
$T_6/Z_3$. 
Therefore, after the $Z_3$ moding, only one pair of electric and magnetic charges
survives, consistently with the fact that, as already pointed out at the beginning of 
this section, only one vector field survives to the projection in the low energy 
effective theory, namely the graviphoton.
The fact that the Dirac quantization condition still holds, like in the 
$T_6$ case, is due to the fact that (\ref{orbch}) can be seen as the superposition of 
$3$ pairs of $3$-branes on $T_6$, with relative angles $\theta_a + 2 \pi z_a$ instead 
of $\theta_a$. For each pair (\ref{torch}) holds and so for their sum, that is 
(\ref{orbch}).
 
Summarizing, the net effect of the wrapping of the 3-brane on $T_6/Z_3$ is therefore 
to obtain a 1-parameter family of dyons (rather than 4 as for $T_6$) whose effective 
couplings depend only on one combination of the relative angles between the whole 3-branes.  

It is interesting to remark that the $Z_3$ projection, which reduces the 4 independent 
gauge fields to 1, is also responsible for the decoupling of the scalars fields from the 
3-brane, as seen in ref. \cite{hins1}. Thus, the 3-brane wrapped on $T_6/Z_3$ looks like an
extremal R-N configuration, being a source of Gravity and Maxwell field only, the mass and
the dyonic charge being equal in suitable units, i.e. $M^2 = e^2 + g^2$.

\begin{center}{\bf ACKNOWLEDGEMENTS}\end{center}

The authors would like to thank P. Fr\'e and M. Trigiante for illuminating 
discussions on the Supergravity interpretation of the results, E. Gava and 
K.S. Narain for valuable enlightments on the orbifold construction, C. N\'u\~nez for 
early discussions on the topics of this paper. M. B. also thanks A. Lerda, R. Russo and 
G. Bonelli for interesting discussions. 
This work has been partially supported by EEC contract ERBFMRXCT96-0045.


\end{document}